\RequirePackage{amsmath}
\documentclass{llncs}

\usepackage{epstopdf}
\usepackage{graphicx}

\usepackage{setspace}
\usepackage{color}   %for hilite command
\usepackage[utf8x]{inputenc}
\usepackage[QX,T1]{fontenc} 
\usepackage{xspace}
\usepackage{wrapfig}
\usepackage{balance}
\usepackage{tikz}
\usepackage{url}
\usepackage{mdwlist}
\usepackage{subfigure}
\usepackage{mdwlist}
\usepackage{bm}
\usepackage{booktabs}
\usepackage{multirow}
\usepackage{hhline}
\usepackage[ruled,linesnumbered]{algorithm2e}
\usepackage{algorithmic}

\usepackage[font=small,labelfont=bf,skip=0pt]{caption}
\usepackage{rotating}
\usepackage{array}
 \usepackage{caption}
\usepackage{enumitem}

\usepackage{adjustbox}

\usepackage[linesnumbered,ruled]{algorithm2e}
\usepackage{hyperref}

\newcommand{\hide}[1]{}

\newcommand{\reminder}[1]{{\textcolor{red}{#1}}}

\newcommand{\OffComm}{OffensiveCommunity\xspace}
\newcommand{\OffCommShort}{OffensiveComm.\xspace}
\newcommand{\HTS}{HackThisSite\xspace}
\newcommand{\WS}{WildersSecurity\xspace}

\newcommand{\WSShort}{WildersSec.\xspace}
\newcommand{\ethical}{EthicalHackers\xspace}
\newcommand{\darkode}{Darkode\xspace}

\newcommand{\myalg}{RIPEx\xspace}

\newcommand{\seedMethod}{Initialization via domain adaptation\xspace}

\newcommand{\WordsFreq}{\textit{TextInfo}\xspace}
\newcommand{\IPP}{\textit{DecimalVal}\xspace}
\newcommand{\Mixed}{\textit{Mixed}\xspace}

\newcommand{\extUserWord}{\textit{ContextInfo}\xspace}
\newcommand{\extUserWordlong}{Contextual Information\xspace}

\newcommand{\PostTextlong}{Text information of the post\xspace}   %% FIRST CAPITAL

\newcommand{\postText}{\textit{PostText}\xspace}

\newcommand{\mfal}[1]{{\bf{\textcolor{blue}{Michalis:}}{\textcolor{red}{#1}}}}

\newcommand{\idenp}{Identification\xspace}
\newcommand{\classp}{Characterization\xspace}

\newcommand{\crosstrain}{cross-training\xspace} %% WE want this to imply learning from another forum, when we lack seed data

\newcommand{\cseed}{Cross-Seeding\xspace} 

\newcommand{\cporting}{Basic\xspace} %% name for when we just tranfser the classifier

\newcommand{\kwords}{$W$\xspace}
\newcommand{\kwordsDef}{Word-Range\xspace}

\begin{document}
\title{\myalg: Extracting malicious IP addresses from security forums using cross-forum learning}

\author{Joobin Gharibshah  \and
Evangelos E. Papalexakis \and Michalis Faloutsos}
\institute{
University of California - Riverside, CA\\
900 University Ave, Riverside, California 92557\\
\email{jghar002,epapalex,michalis@cs.ucr.edu}}

\maketitle

\vspace{-0.4cm}
\begin{abstract}

Is it possible to extract malicious IP addresses reported in security forums
in an automatic way?
This is the question at the heart of our work.
We focus on security forums, where security professionals and hackers share knowledge and information, and often report misbehaving IP addresses.
So far, there have only been a few efforts to extract information from such security forums.  We propose \myalg, a systematic approach to identify and label IP addresses in security forums by utilizing a cross-forum learning method. In more detail, the challenge is twofold: (a) identifying IP addresses from other numerical entities, such as software version numbers, and (b) classifying the IP address as benign or malicious. 
We propose an integrated solution that tackles both these problems. 
A novelty of our approach is that it does not require training data for 
each new forum. Our approach does knowledge transfer across forums: we use 
a classifier from our source forums to identify seed information for training a classifier on the target forum.
%using transfer the learned knowledge between forums.
We evaluate our method using data collected from five security forums 
with a total of 31K users and 542K posts. First, \myalg can distinguish IP address from other numeric expressions with 95\% precision and above 93\% recall on average. Second, \myalg identifies malicious IP addresses with an average precision of 88\% and over 78\% recall, using our cross-forum learning.
%\mfal{Third, we are able to extract XX more malicious IP addresses compared to VirusTotal database. }  
% 15K IP addresses in five forums - 
Our work is a first step towards harnessing the  wealth of useful information that can be found in security forums.

\end{abstract}

\vspace{-0.4cm}

\section{Introduction}
\label{sec:intro}

\begin{figure}
 \centering
 %\begin{center}
\includegraphics[width=1\textwidth]{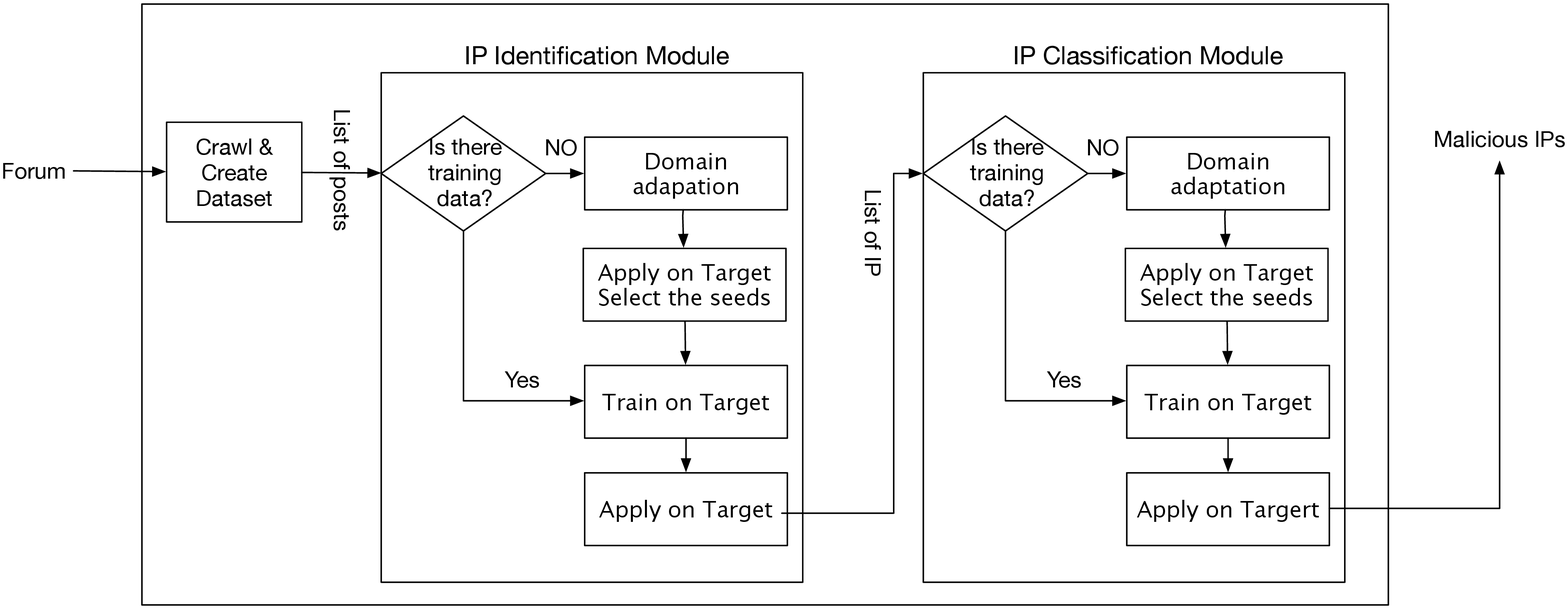}
 %\end{center}
 \vspace{0.01cm}
 \caption{The overview of key modules of our approach (RIPEx): (a) collecting data, (b) IP \idenp, and (c) IP \classp. In both classification stages, we use our \cseed approach that  in order to generate  seed information for training a classifier for a new forum.\label{fig:sysModel}}
\vspace{-0.75cm}
\end{figure}

The overarching goal of this work is to harness the user generated content in forums, especially security forums. More specifically, we focus here on collecting malicious IP addresses, which are often reported at such forums. 
We use the term security forums to 
refer to discussion forums with a focus on 
security, system administration, and in general systems-related discussions.
In these forums, security professionals, hobbyists, and hackers  identify issues, discuss solutions, 
and in general exchange information.

We provide a few examples of the types of discussions that take place in these forums that could involve IP addresses, which is our focus. Posts could talk about a benign IP address, say in configuration files, 
%a server that may not be responding or exhibiting unusual delays. 
 as in the post: {\it "[T]his thing in my hosts file:   64.91.255.87 ... [is] it  correct?"}.
At the same time, posts could also  report compromised or malicious IP addresses,
as in the post: {\it "My browser homepage has been hijacked to  {\small http://69.50.191.51/2484/}"}.
Our goal is to automatically distinguish between
the two and provide a new source of information
for malicious IP addresses directly from the affected
individuals.

The problem that we address here is to find all the IP addresses that are being reported as malicious in a forum.
In other words, the input is all the posts in a forum and the expected
output is a list of malicious IP addresses.
As with any classification problem, one would like to achieve both high precision and recall. Precision
represents the percentage of the correctly labeled over all addresses labeled malicious. Recall is
the percentage of malicious addresses that we find among all malicious addresses reported in forums.
It turns out that this is a two-step problem.
First, we need to solve the {\bf IP \idenp} problem:  distinguishing IP addresses
from other numerical
 entities, such as a software version.
Second, we need to solve the {\bf IP \classp} problem: characterizing IP address
as malicious or benign.
The extent of the \idenp problem caught us by surprise:
we find 1820 non-address dot-decimals, as we show in table \ref{tab:forums}.
%the data told us otherwise.

% ------ Michalis stopped here I will resume in a bit ----

%\mfal{Previous work}
There is  limited work on extracting information from security forums,
and even less work on extracting malicious IP addresses. 
% Overall, there has not been a comprehensive systematic approach, which could 
% address this problem at a massive scale.
We can group prior work in the following categories.
 First, recent works %, including our prior work,
 study the number of malicious IP addresses in 
 forums,
but without providing the comprehensive and systematic solution that we propose here~\cite{Frank2016}. 
Second, there are recent efforts that extract
other types of information from security forums,
related to the black market of hacking services and tools \cite{Portnoff2017}, or
  the behavior and roles  of their users~\cite{Holt2012,Abbasi2014}.
 Third, other works focus on analyzing structured sources, such as security reports   and vulnerability databases~\cite{BridgesJIG13,Jones2015}.
  We discuss related work in section~\ref{sec:related}.
 
 There is a wealth of information that
 can be extracted from security forums, which motivates this research direction.
 Earlier work suggests that there is
 close to four times more malicious IP addresses in forums
 compared to established databases of such IP addresses~\cite{Joobin2017}.
 At the same time, there are tens of thousands of IP addresses in the forums, as we will see later.
 Interestingly, not all of the reported IP addresses
 are malicious, which makes the classification necessary.
 
%  However, there exists very little work \reminder{V: again, cite!,Done} on extracting entities from unstructured and informal language, usually present in discussion forums; more specifically, none of the existing works focuses on extracting IP addresses\cite{Portnoff2017}.
% %over unstructured and informal language of the forums there are a few research to extract entities while none of them touch IPs as a point of interest.

%Network analysts usually need to check reports of blacklist repositories or virus detection services like \textit{virustotal.com} to obtain information regarding the maliciousness of and IP address. There are a large number of these repositories that would be needed to be checked in order to identify the malicious IP addresses. However, it would be possible to realize  the maliciousness identity of an IP by looking into users discussions about them in security forums. For instance,  people can run into a problem with an IP address and discuss about it in the forum before it is blacklisted and be reported in malicious repositories. So analyzing the appearing IP address let us know about them before they are reported by others. 

We propose \myalg \footnote{
In the spirit of  double-blind reviewing, we withhold the
explanation of the acronym.
%\myalg stands for \myalgFull.
}, a comprehensive, automated solution that can detect malicious IP addresses reported in security forums. As its key novelty, our approach minimizes the need for human intervention. 
First, once initialized with a small number of
security forums, it does not require
additional training data to mine new forums.
Second, it addresses both the \idenp and \classp problems.
Third, our approach is systematic and readily deployable.
We are not aware of prior work claiming these 
three properties,
as we  discuss in section~\ref{sec:related}.
The overview of our approach is shown in figure~\ref{fig:sysModel}.

The key technical novelty  is  that we
propose {\bf \cseed}, a method to 
%initialization via domain adaptation
conduct a multi-step knowledge transfer across forums.
  We use this approach for both  classification problems,
  when we have no training data for a new forum.
 With \cseed, we create training data for the new forum in the process depicted in figure~\ref{fig:sysModel}.
We use a classifier based on the current forums to identify seed information in the new forum.
We then use this seed information to train
a classifier for the new forum.
This forum-specific classifier performs much better
than if we have used the classifier of the current forums on the new forum. 
We refer to this latter knowledge transfer approach 
as {\bf \cporting}. 

We evaluate our approach using five security forums with 
a total of 31K users and 542K posts spanning a period of roughly six years.
% We establish our ground-truth using VirusTotal, which is a site that aggregates malicious IP addresses from many different sources and some manual inspection.
Our results can be summarized into the following points.

{\bf a. \idenp:  98\% precision with training data  per forum.}
We develop a supervised learning algorithm for 
solving the \idenp problem 
in the case where we have training data for the target forum.
Our approach exhibits 98\% precision and 96\% recall on average across all our sites.
% Interestingly,  the numerical value of the IP address is an important feature: addresses have higher numerical values than non-IP numerical expressions. 
%such as product versions as we discuss later.
% \reminder{isn't it better to mention it on Evaluation section? and mix of two is even better (contextual+ip parts)}

{\bf b. \idenp: 95\% precision with \cseed.} 
We show that our \cseed approach is effective in
transferring the knowledge between forums.
Using the \WS forum as source, we observe an average of 95\% precision and 93\% recall in the other forums.

% What is improvement over "non intelligent" \crosstrain? On average it is 8\% improvement. \mfal{on prevision?}

{\bf c. \classp: 93\% precision with training data per forum}. We develop  a supervised learning algorithm for 
solving the \classp problem 
assuming we have training data for the target forum. Our classifier achieves 
93\% precision and 92\% recall on average across our forums. 

{\bf d. \classp: 88\% precision on average with \cseed data.} 
We show that our \cseed approach by using \OffComm forum as source can provide
88\% precision and 82\% recall on average.
% 

%What is improvement over "non intelligent" \crosstrain?

{\bf e. \cseed  outperforms \cporting.} We show
that \cseed is important, as it increases the precision by 28\% and recall by 16\% on average in the \classp problem, and the precision by 8\% and recall by 7\% on average in the \idenp
problem. 
% for 5 WS, and 6 OffensiveComm

{\bf f. Using more source forums improves the \cseed performance.} We show that, by adding a second source forum, we can improve the precision by 13\% on average over the remaining three forums. 
%compared to using one prior forum.

Our work suggests that there is a wealth of information
that we  find in security forums and offers a systematic approach to do so.

%\mfal{---------- NOTES -------------}
%\mfal{
%We need to use consistently one expression: domain adaptation and transfer learning or initialization via domain adaptation... ASK VAGELIS.
%}

\hide{
---------- NOTES -------------

\mfal{We need to define three names:}

\crosstrain: this to imply the problem of needing to learning from another forum, when we lack seed data 

Two solutions to this problem:

\cporting: name for when we just transfer the classifier - "Naive solution"

\cseed: Cross-Seeding: to imply our approach, use external classifier to create a seed. "Our solution".

-------------

\mfal{Richness of information: If we don't results here on wide applicatoin of RIPEx, we could refer to our ASONAM paper.}

How to present the story:
a. as a framework up front altogether

b. or build it slowly with a view to practical appliation: IP identification, learning with seed, learning without seed

---------- Old parts -----------

To this end, we propose a supervised learning method, leveraging simple and interpretable context-based features, to identify IP addresses from discussion forums, and label them as malicious or benign.

For supervised learning to work successfully, we need high-quality labeled data, which may be expensive and time-consuming to come by. Typically, obtaining reliable ground-truth, entails a large amount of manual labeling, usually done by annotators familiar with security-related topics, and doing so can be time-consuming and expensive. What if we only have a budget that allows us to reliably label data for a specific security forum. How can we re-use the knowledge learned by this tedious labeling process, in order to detect malicious IPs in different forums, for which we have no labels, and perhaps requiring labeling would defeat the purpose of the detection, since detecting an IP is a very {\em time-sensitive} task. To that end, we propose an effective domain adaptation technique, which leverages seeding, in order to transfer knowledge between a forum for  which we already have ground truth, to new and unseen forums \reminder{V: maybe we also need to say that those forums are not TOO similar with each other}. Our proposed framework is presented in figure \ref{fig:sysModel}.
%All the supervised learning task need label data. Which is expensive and rare in social network analysis. In the case of having no  human annotators to label the data or validate the given label, we can use other resources and transfer possible knowledge to mitigate the labeling effort and automate the identification and labeling task. To this end, a simple domain adaptation method along with a new seeding method has been proposed to transfer knowledge form other datasets in to the target domain. So highly confident labels assigned by a source domain trained model are used in the target domain to provide the ground truth and training data to build a model on the target domain. The propose model is presented in the figure \ref{fig:sysModel}. 
This model gets the forums \reminder{"forum's posts" if we exclude the crawling} as an input and in two modules identify and mark them as malicious an benign. In each module based on the existence of the labeled data either direction learning (with labeled data) or \FeedMethod (with transferred knowledge) are being taken.

%So far, most of the work on social media analysis focuses on social platforms, such as twitter and Facebook, and very little work focuses on security forums. In fact many efforts  focus on addressing security problems using knowledge obtained from the web, as well as, social and information networks, these efforts are mainly focused on analyzing structured sources (e.g., \cite{Iannacone2015}).  
% However, studies assessing the usefulness of (unstructured) information in online forums have only recently appeared (e.g., \cite{Samtani2015}). 
%5 These studies are mostly exploratory in that they provide evidence of the usefulness of the data in the forums, but do not provide a systematic 
 %methodology or ready-to-use tools, which is the goal of our work.
 %We discuss existing literature in more detail later in this section

Our results can be summarized into the following points:

a) We propose a supervised learning method to exclude the numerical patterns in the text which are not actual IP address in average with 98\% accuracy on the balanced datasets.

b) We develop a supervised learning method to label malicious IP based on contextual features in average with 93\% accruacy on balanced datasets in five of our forums. 

c) We propose a \seedMethod to identify IP address and label them as malicious across dataset by having other datasets as the resource for building the models in order to identify IP addresses in average with 96\% precision and label them as malicious and benign in average with 88\% precision. 

%d) We show that our method can find roughly “2-5” times more malicious IP addresses compared to VirusTotal and we find 51-73\% (average over all forums) of the malicious IPs earlier than VTotal among the jointly-found malicious IPs.
%[ Do average over all forums:  4 times more IPs addresses and from the jointly detected addresses, ~62\% are detected earlier with ~45\% of these addresses earlier by at least three months!)

%e) We study the spatiotemporal properties of the malicious IPs we found. We find that: (i)  In 2010 and 2014 InferIP in \WS reported 6 and 7 more malicious IPs than VT (spikes on the time series),  (ii) 80\% of malicious IPs seem to be in 3 countries (United States, United Kingdom and Australia).

Regular expression is a suggested method to extract IP from text while it might ends up to high false positive rates by grabbing all dot-decimal expressions like IP addresses as real IP addresses. For example, in a post discussing about software versions like \textit{"Title:spywareblaster wont install Post: ChatSpace Java Client 2.1.5.95 ..." } the dot-decimal expression \textit{"2.1.5.95"} looks like a valid IP address but over the context it was presented as another entity .i.e software version. 

Second, after extracting IP addresses from the text,  the next challenge is to classify the IP addresses into malicious and benign, based on their context.
Typically, identification of malicious IPs is done manually by network analysts, who need to compare a particular IP against reports and blacklist repositories, such as  \textit{virustotal.com}, in order to verify their maliciousness. There exist a large number of such repositories, and worse yet, they may not be complete. In this paper, we propose to leverage the {\em context} of a particular IP address mention, in order to extract useful information regarding its maliciousness. For instance, we may be able to extract useful information by looking into the discussion surrounding an IP address: very frequently, problematic IPs are being discussed in security forums, long before they are blacklisted and appear in an online repository. Thus, analyzing the textual context around an IP address can reveal critical information regarding its maliciousness.

}
\section{Our Forums and Datasets}
\label{sec:data}

We have collected data from five different forums,
which  cover a wide spectrum of
interests and intended audiences.
We present basic statistics of our forums in Table~\ref{tab:forums} and we highlight the
differences of their respective communities. 

{\bf Our semi-automated crawling tool.} We have developed an efficient and customizable  python-based crawler, 
which can be used to crawl online forums,
and it could be of independent interest.
% ANONYMIAZATION:
% An earlier version of the crawler was described in our prior work~\cite{Joobin2017}.
To crawl a new forum, our tool requires a 
configuration file that describes the structure of the forum. 
Leveraging our current configuration files, 
the task of crawling a new forum is simplified significantly. Due to space limitations, we do not provide further details. Following are the descriptions of collected forums.
% As we have observed the activity and posting on all these forum they have different attitude in the security domain.

\let\labelitemi\labelitemii
\begin{itemize}

 \item {\bf\WS (WS)} seems to attract system administrator types
and focuses on defensive security: how one can manage and protect one's system. Its topics include anti-virus software, best practices, and  new vulnerabilities and its users seem professional and eloquent.  %They definitely seem like the older, wiser, not as fast paced, more calm group where other forums seem to be more filled with youth and energy, lack of guidance, and more malicious intent. 

  \item {\bf \OffComm (OC)} seems to be on the fringes 
 of legality. As the name suggests, the forum focuses on breaking into systems: it  provides step by step instructions, and  advertises hacking tools  and services.

 \item {\bf \HTS (HT)} seems to be in between these extremes represented by the first two forums. 
% They are not as malicious as offensive community, but are not as defensively oriented as \WS.
For example, there discussions and competitions on   hacking challenges, but it does not act as openly as a black market of illegal services and tools compared to \OffComm. 

 \item {\bf \ethical (EH)} seems to consist mostly of  ``white hat" hackers, as its name suggests. The users 
discuss hacking techniques, but they seem to have a strict moral code. 

 \item {\bf \darkode (DK)} is a forum on the dark web that has been taken down by the FBI in July 2015. The site  was a black market for malicious tools and services similar to \OffComm. % The software being traded had to be some sophisticated malware for it to be targeted by the FBI.
\end{itemize}

Our goal is to identify and report IP addresses that the forum readers report as malicious. We currently do
not assess whether the author of the post is right, though
the partial overlap with blacklisted IPs indicates so. 
We leave for future work to detect misguided reports of IP addresses.

\begin{table}[t]
 \centering 
%\label{tab:forums}
\begin{adjustbox}{width=.98\textwidth}
\small
 \begin{tabular}{|l|r|r|r|r|r|} 
 \hline
  &\WSShort & \OffCommShort & \HTS & \ethical & \darkode \\ 
 \hline
 Posts &302710  & 25538  & 84125  &54176 & 75491 \\ 
 \hline
 Threads & 28661 & 3542 & 8504 & 8745 & 7563\\ 
 \hline
 Users & 14836 & 5549 & 5904 & 2970 & 2400\\ 
 \hline 
 Dot-decimal & 4325 & 7850 & 1486 & 1591 & 1097\\
 \hline
 IP found & 3891 & 6734 & 1231 & 1330 & 1082\\ 
 \hline
\end{tabular}
\end{adjustbox}
 \caption {The basic statistics of our forums \label{tab:forums}}
 \vspace{-1cm}
\end{table}

{\bf Determining the ground-truth.}
For both of the problems we address here, there are
no well-established benchmarks and labeled datasets.
To train and validate our approach, we had to rely on
external databases and some manual labelling. 
For the \idenp problem, we could not find any external sources of information and benchmarks. To establish
our ground-truth, 
we selected dot-decimal expressions uniformly randomly,
and we used four different individuals for the labelling. 
To ensure testing fairness, we opted for balanced datasets,  which led us to a corpus 
of 3200 labeled entries across all our forums.

For the \classp problem, we make use of the VirusTotal site
which maintains a database of malicious IP addresses by aggregating information from many other such databases. 
We also provide a second level of validation 
via manual inspection.

We create the ground truth by uniformly randomly selecting and assessing IP addresses from our forums.
If VirusTotal and the manual inspection give it the same label, we add the addresses into our ground-truth.
Finally, we again ensure that we create balanced sets
for training and testing to ensure proper training and testing.

% If VirusTotal and the 
%  manual inspection  not malicious, we would add it in our ground-truth set of non-malicious addresses.
% \mfal{@Joobin: check that this is what we do!} \reminder{Done, I have changed the text}

% Finally, we did an additional pass to ensure 
%  the ground-truth posts met some specifications
%  that would make them suitable for training and testing.
%  For example, we wanted our posts to have textual context, since a post with minimal text would not be helpful for training.

% Thereafter, the human narrator consider the marked IPs by {\tt VirusTotal} and exclude IPs from malicious list if they appeared in a post with not clues to be malicious. 
% Similarly for the benign list, if the IP appears in a post with a malicious content it was excluded form that list as well.

\section{Overview of  \myalg}
\label{sec:ourApproach}

We represent the key components of our approach
in addressing the \idenp and \classp problems.
To avoid repetitions,
we present at the end the \cseed approach,
which we use in our solution to both problems.

\subsection{The IP \idenp module}
\label{sec:ipDetection}

We describe our proposed method to identify IP addresses in the forum. 

{\bf The IP address format.}  The vast majority of IP addresses in the forums follow the \textit{IPv4} dot-decimal format, which  consists of 4 decimal numbers  in the range [0-255] separated by dots.
We can formally represent the dot-decimal  notation as follows: \textit{IPv4} $ [x_1.x_2.x_3.x_4]$ with $ x_i \in  [0-225]$, for $i = 1,2,3,4$.  
Note that the newer \textit{IPv6} addresses consists of eight groups of four hexadecimal digits, and our algorithms could easily extend to this format as well.
Interestingly, we found a negligible number of \textit{IPv6} addresses, and we opted to not
focus on \textit{IPv6} addresses here.
For example, in \WS forum, we find 3891 \textit{IPv4} addresses and only 56 \textit{IPv6} addresses.
At such small numbers, it is difficult to train and test a classifier.
% It is also possible that on the text IP addresses are noted in hexadecimal, octal or binary representation which we neglect them. Since we could not extract noticeable number of them in our point of interest domain so it is out of scope of this research. 
Thus,  
% we  only discuss \textit{IPv4} addresses \reminder{, simply called IP address,} in the rest of this paper.
for the rest of this paper,
IP address refers to \textit{IPv4} addresses.

{\bf The challenge: the dot-decimal format is not enough.}
If IP addresses were the only numerical expressions in the forums with this format, the \idenp problem could have been easily solved with straightforward text processing 
 and Named-Entity Recognition (NER) tools, such as the Stanford NER models~\cite{Finkel2005}.
However, there is a non-trivial number of other numerical expressions, which can be misclassified as addresses. 
For example,  we quote a real post: \textit{"factory reset brings me to the Clockworkmod 2.25.100.15 recovery menu"}.
where the structure
 \textit{2.25.100.15}  refers to the version of Android app \textit{"Clockworkmod"}.

% Since other similar dot-decimal structure like "software version" can mistakenly be captured as IP addresses by regular expression. Moreover, there is no dataset and ready-made model considering IP address as an entity in the context of forums in which people use non-formal and written language with variety of internet lingoes. 
% For example, people address to a dot-decimal structure like \textit{5.3.18.129} as a part of a  system log in a post such \textit{"factory reset brings me to the Clockworkmod 2.25.100.15 recovery menu"}. Apparently the entity (\textit{5.3.18.129}) has a similar pattern to an IP Address, therefore it is unidentifiable by only using regular expressions. 

To this end, we propose 
a method to solve the IP \idenp
problem, a supervised learning algorithm.
We first identify the features of interest as we discuss below.
We then train a classifier using the Logistic Regression method gives the best results among the several methods using 10-fold cross validation on our ground-truth as we decribed in the previous section.

{\bf Feature selection.}
We use three  sets of features in our classification.

{\bf a. Contextual information: \WordsFreq.} Inspired by how a human would determine the answer, we  focus on the
 words surrounding the dot-decimal structure.  
 For example, the words \textit{"server"} or \textit{"address"} suggests that the dot-decimal is an address, while the words \textit{"version"} or a software name, like \textit{"Firefox"} suggests the opposite. 
 At the same time, 
 we wanted to focus on  words close to
 the dot-decimal structure.
 Therefore, we introduce {\bf \kwordsDef,  \kwords, } to determine the number of surrounding words before and after the dot-decimal structure that we want to consider in our classification.
 We use TF-IDF~\cite{Ramos2003} to normalize the frequency
 of a word to better estimate its discriminatory value.

{\bf b. The numerical values of the dot-decimal: \IPP.}
We use the numerical value of the four numbers in the 
the dot-decimal structure as features. The rationale is that  non-addresses, such as software versions, tend to have lower numerical values. This insight
was based on our close interaction with the data.
% Note that we introduce

{\bf c. The combined set: \Mixed.} We  combine the two feature sets to create in order to
 leverage their discriminating power.

\vspace{-15pt}
\begin{figure}
    \centering
    \begin{minipage}{0.49\textwidth}
        \centering
        \includegraphics[width=\textwidth]{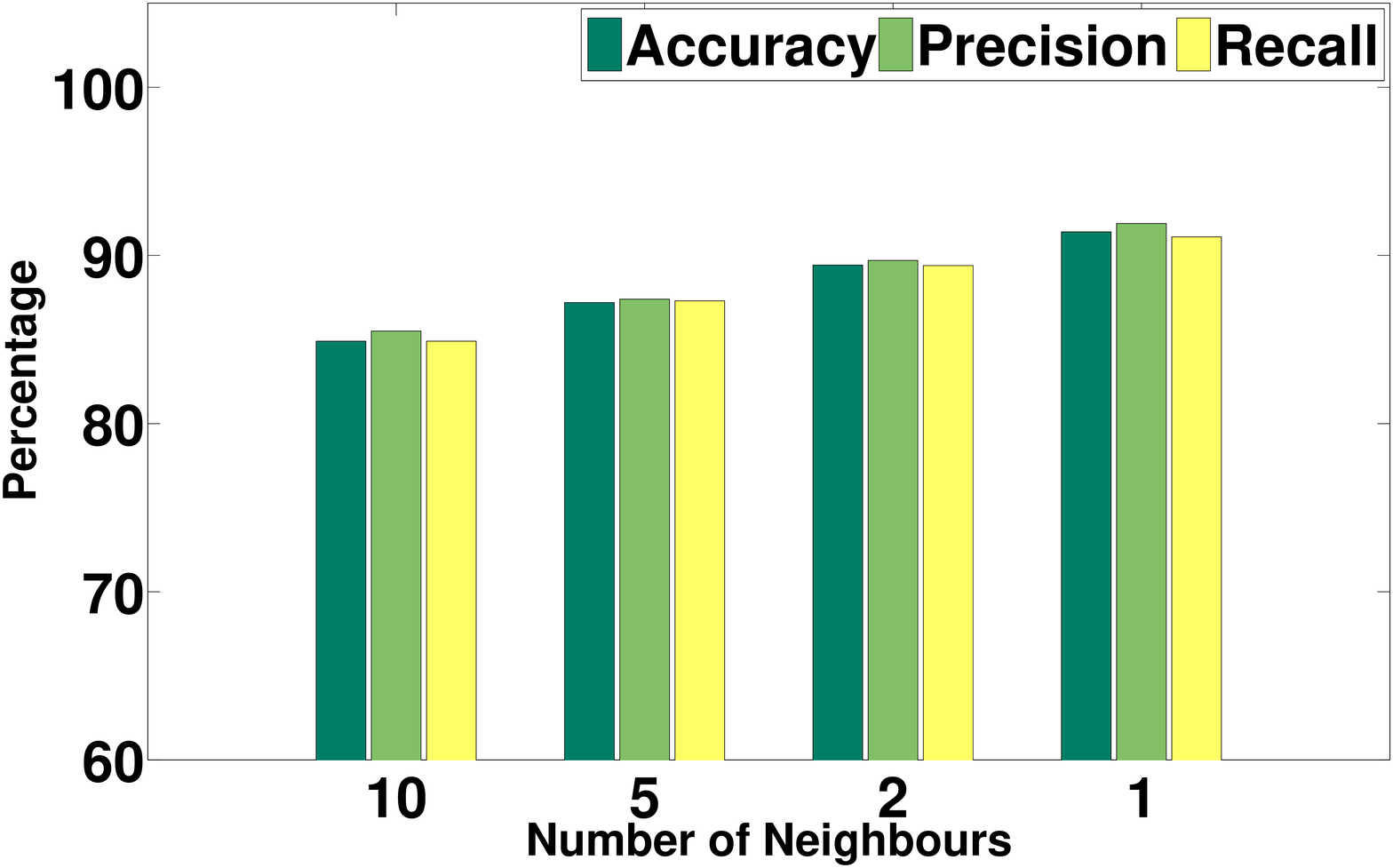} % first figure itself
        \caption{Classification performance versus 
        the number of words \kwordsDef,~\kwords, in \WS. }\label{fig:IPNeighbours}
        \end{minipage}\hfill
    \begin{minipage}{0.49\textwidth}
        \centering
        \includegraphics[width=\textwidth]{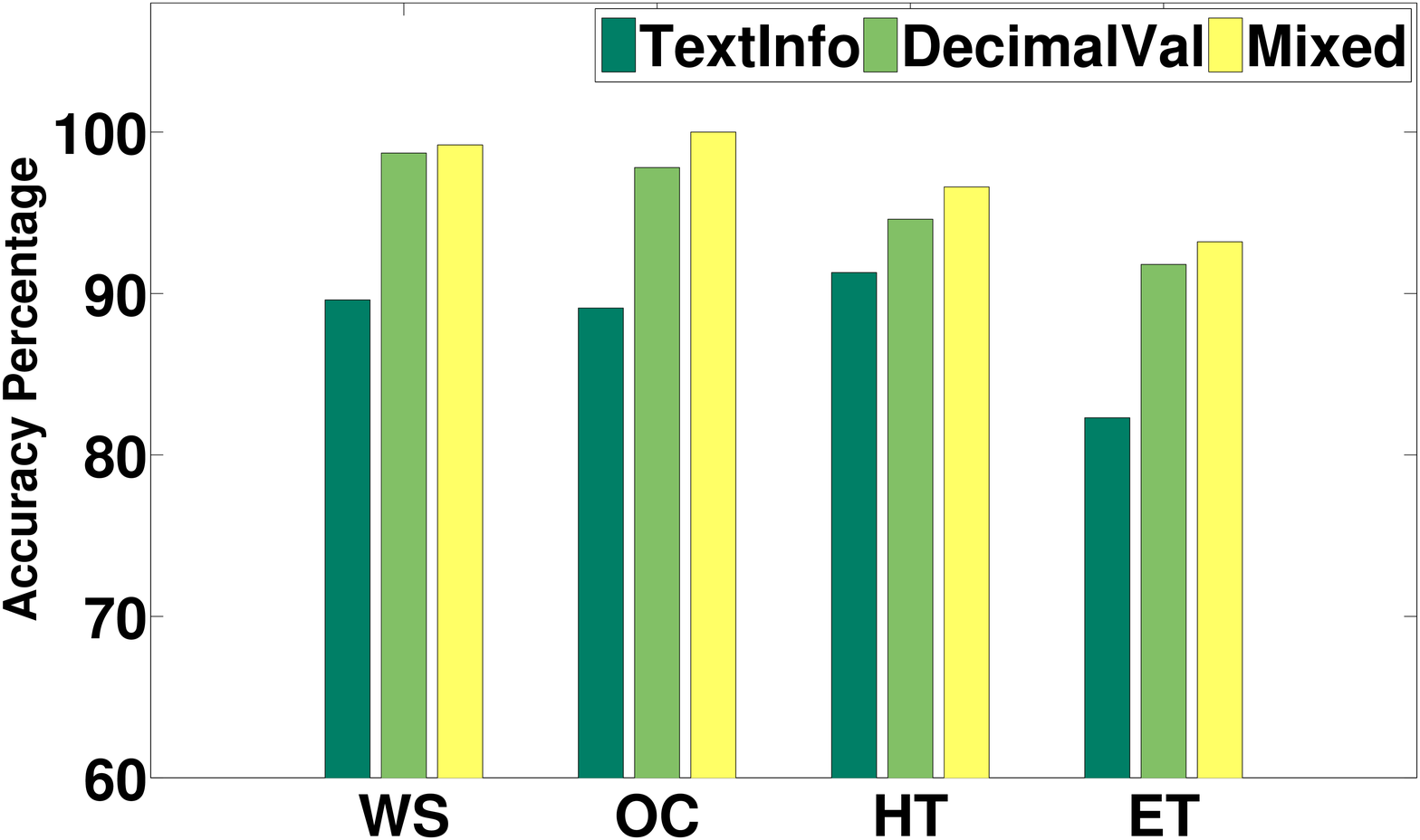} % second figure itself
   \caption{Classification accuracy for different features sets in 10-fold cross validation in four forums.\label{fig:IPFeatures}}
        
    \end{minipage}
    \vspace{-0.5cm}
%    \caption{WildersSec. model compare active and simple cross dataset evaluation to identify IPs}
\end{figure}

% The performance of IP identification module, discussed in  \ref{fig:sysModel}, in the case of having labeled data evaluated by 10-fold cross validation. More precisely, the effect of the different feature sets is evaluated in the \classp in two perspectives.

{\bf Determining the right number of context words, \kwordsDef.}
We wanted to identify the best value of parameter
\kwordsDef for our classification.
In figure~\ref{fig:IPNeighbours}, we plot the 
classification accuracy, precision and recall, as we vary \kwordsDef, \kwords$=  1,2,5~and~10$, for the \WS forum 
and using only the \WordsFreq.
% \mfal{@Joobin: *** in the future let's check this for Mixed!}
We see that using one to two words gives better results compared to using five and ten words.
The explanation to this counter-intuitive result
is that considering more words includes
text that is not relevant for inferring the nature of
a dot-decimal, which we verified manually.

% \mfal{@Joobin: Did you try \kwords when using ONLY Words, or you varied kwords in the combined solution as well? **** We should explain what we did, and ideally we should try it in the combined Mixed solution!} 
% \reminder{for determining the right number of words we have uses only \kwords} 

% DONT DELETE THIS LINE \mfal{@Joobin: Did you try \kwords when using ONLY Words, or you varied kwords in the combined solution as well? **** We should explain what we did, and ideally we should try it in the combined Mixed solution!}

% We show indicative results with the \WS forum. it shows that the fewer number of words around the numerical structure lead in to the higher accuracy, similar results are gained for the other models. 

{\bf Using numerical values \IPP improves the performance significantly.}
In Figure \ref{fig:IPFeatures}, we plot the  classification accuracy of different features sets.
Recall that we are not able to include \darkode forum
due to its limited number of non-IP dot-decimal expressions,
as we saw in \ref{sec:ipDetection}.
We see that using \IPP  features alone, we can get 94\% overall accuracy  and
using both \IPP and \WordsFreq, we get 
98\% overall accuracy  across our forums.
Focusing on the  IP address  class,
we see a 
an average precision of 95\% 
using only \IPP and, 
 98\% using both \IPP and \WordsFreq.

% Second, different feature set proposed in the section \ref{sec:ipDetection} is evaluated. Showing in figure \ref{fig:IPFeatures}, interestingly,  it turns out that the numerical value of the IP address is an important feature. addresses have higher numerical values than non-IP numerical structures, such as product versions. 

% \mfal{If we show numbers in accuracy, let's report those numbers too, and mention that we use balanced training and testing sets, but we can further report the precision and recall for IP addresses.}
% \reminder{Note: the corresponding plot is in accuracy figure \ref{fig:IPFeatures}}

% --- Michalis stopped here ---

% We use  use contextual features from the post
% to identify IP addresses.

% Basically to reduce the search space to identify IP addresses first we apply regular expression to grab all possible IP addresses with the specific dot-decimal pattern. Afterwards, we extract features for the dot-decimal grabbed expressions.
% Inspired by  human narrators' inference for this identification we want to extract informative features. A human reader 
% by considering the dot-decimal structure itself and it's surroundings text, likewise our narrators, can identify IP addresses easily. In this fashion, we are using two set of features in this model similarly.

\vspace{-0.2cm} 
\subsection{The IP \classp module}
\label{sec:MalIP}
\vspace{-0.1cm} 

We develop a supervised learning algorithm to characterize IP addresses.
Here, we assume 
that we have labeled data, and we discuss how we handle the absence of ground truth in section
 \ref{sec:activeLearning}.
 We first identify the appropriate set of features which we discuss below.
We then train a classifier and
% and evaluate it by 10-fold cross validation method.
find that the Logistic Regression method gives the best results among several methods that we evaluated.
Due to space limitations, we show a subset of our results.

\begin{figure}[t]
 \centering
\includegraphics[width=0.5\textwidth]{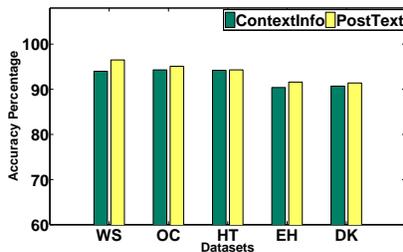}
 \caption{\classp: The effect of the features set on the classification accuracy with balanced testing data.}\label{fig:malFeaturesSets}
 \vspace{-0.5cm}
\end{figure}

{\bf Features sets for the \classp problem.}
We consider and evaluate three  sets of features in our classification.

{\bf a. \PostTextlong: \postText.} 
We use the words and their frequency of appearance in the post. Here, we  use the TF-IDF technique~\cite{Ramos2003}
again to better estimate the
discriminatory value  of a word by considering its 
overall frequency.
% Here, we decided not to use a proximity-based limit to focus on words around the IP address.
% Our reasoning is as follows: 
% (a) we found that getting the sentiment of the post 
% was better obtained by considering the whole post,
% and, 
% (b)  the classification performance was pretty good.
In the future, we intend to experiment
with sophisticated Natural Language Processing models
for analyzing the intent of a post.

{\bf b. The \extUserWordlong set: \extUserWord.} 
We consider an extended feature set that includes
both the \postText features, but also features of
the author of the post. These features
capture the behaviour of the author, including
frequency of posting, average post length etc. 
These features were introduced by
  earlier work~\cite{Joobin2017}, with
the rationale that  profiling  the author
of a post can help us infer their intention and role
and thus, improve the classification.

% {\bf c. Co-clustering feature set: \Cocluster.}
% We consider the behavioral information of the users,
% such as 

% Previously we introduced a co-clustering based method to  to extract the latent features and classify the IP addresses. Apply the co-clustering method on the \Mixed feature set generate the last feature set in this study, called \Cocluster.  

{\bf \classp: 93\% precision with training data.} 
We assess the performance of the \classp classifier using the   set of features above and by 
using the labeled data of each forum.
We evaluate the performance using 10-fold cross validation.
In figure \ref{fig:malFeaturesSets}, we show the accuracy of classification.

We can achieve 
93\% precision and 92\% recall on average across all the forums. 
The results are shown in figure~\ref{fig:malFeaturesSets},
where we report the results using the accuracy across both
classes, given that we have balanced training datasets.

{\bf  Selecting the \postText feature set.}
We see that, by using \postText features on their own, we obtain slightly better results.
\postText feature achieves 94\% accuracy on average, 
while using the \extUserWord results in 92\% accuracy on average across all forums.
Furthermore, text-based only features have one more key advantage: they can transfer between domains in
a straightforward way.
Therefore, we use the \postText features 
in the rest of the paper.

% Since we are willing to use a feature set which can be transferable into other dataset in section \ref{sec:activeLearning}, therefore we select \WordsFreq in \classp problem.

% \mfal{Should we leave this here or move it to next section?}

% \reminder{I have moved it here but there is problem with the figure is in accuracy and we discuss here about precision and recall. Should we explain about accuracy here as well?}

% \mfal{We should revisit this. Should we move "training data" to algorithm section or Evaluation?} 

%\begin{figure}[h]
% \centering
%\includegraphics[width=0.5\linewidth]{CompareDatasets}
 %\caption{Comparing different feature set in classifying maliciousness of the IPs}\label{fig:malFeaturesSets}
%\end{figure}
%
%\begin{figure}[h]
% \centering
%\includegraphics[width=0.5\linewidth]{WilderPrecisionActive.eps}
% \caption{Wilder precision}\label{fig:adfadf}
%\end{figure}

\vspace{-0.3cm}
\subsection{Transfer Learning with \cseed}
%\section{Transfer Learning by domain adaptation}
\label{sec:activeLearning}

\vspace{-0.1cm}

In both classification problems, we face the following conundrum:

a. the classification efficiency is better when 
the classifier is trained with forum-specific ground-truth, but, 

b. requiring ground-truth for a new forum will introduce manual intervention, which will limit the practical value of the approach.

We propose to do cross-forum learning by leveraging
transfer learning approaches~\cite{daumeiii2007,pan2010}.
We use the terms {\it source} 
and {\it target} domain
to indicate the two forums with the target forum not
having ground-truth available.
For both classification problems, we consider
two solutions for classifying the target forum:

 {\bf a. \cporting:} We use the classifier from the source forum on the target forum.
 
 {\bf b. \cseed:} We propose an algorithm that
 will help us develop a new classifier
 for the target forum by using the old classifier to 
 create training data as we explain below.

\vspace{-0.2cm} 
\begin{algorithm}
\caption{\cseed:  transfer learning between forums\label{alg:seedalgo}}
%\begin{algorithmic}[1]
%\Procedure{CrossDomain} $(\mathcal{X},\mathcal{Y}):$\
{CrossForum} $(\mathcal{X},\mathcal{Y}):$\

\quad Take the union of the features in forum $\mathcal{X}$ and $\mathcal{Y}$

\quad Apply classifier from $\mathcal{X}$ on  $\mathcal{Y}$ 

\quad Select  the high-confidence instances
to create seed for $\mathcal{Y}$

\quad Train a new classifier on $\mathcal{Y}$ based on the new seeds. 

\quad Apply the new classifier on $\mathcal{Y}$

%\end{algorithmic}
\end{algorithm}
 
 \vspace{-0.4cm}
{\bf Our \cseed approach.}
We propose to create training data for the target forum 
following the four steps below, which are illustrated  in figure~\ref{fig:sysModel} and outlined in 
algorithm~\ref{alg:seedalgo}.
% We use a classifier from the current forums to identify seed information for training a classifier for the new forum.
% We then use the seed information to train
% a classified for the new forum in question.

{\bf a. Domain adaptation.}
The main role of this step is to ensure that the
source classifier can be applied to the target forum.
The main issue in our case is that the feature sets
can vary among forums. 
Recall that, for both classification problems,
we use the frequency of words and these words can vary
among forums.
We adopt an established approach
that works well for text classification~\cite{daumeiii2007}:
we take the union of the feature sets of the source and target forums.
%which is one way of the possible ways to handle this,
The approach seems to work sufficiently well
in our case, as we see later. 

%\reminder{should we cite? \cite{daumeiii2007}}
% unify the target and source features set by mapping them into a high-dimensional domain where we replicate the common and target specific features in the source domain and similarly common and source specific features in to target domain.

{\bf b. Creating seed information for the target forum.}
Having resolved any potential feature disparities,
we can now apply the classifier from the source forum
to the target forum.
% After augmenting feature set on \textit{S} and \textit{T} domain, it is possible to train a traditional learner like logistic regression on \textit{S} and test it on \testit{T} domain.
We create the seeding data by  selecting   instances of the target domain, 
for which the classification confidence is high.
Most classification methods
provide a measure of confidence for each classified instance and we revisit this issue in section~\ref{sec:evaluation}.

{\bf c. Training a new classifier for the target forum.}
Having the seed information, this is now a straightforward
step of training a classifier.

{\bf d. Applying the new classifier on the target forum.}
In this final step, we apply our newly-trained forum-specific classifier on the target forum.

\section{Evaluation of our Approach}
\label{sec:evaluation}

We evaluate our approach focusing on the 
performance of \cseed for both the \idenp
and the \classp problems.

{\bf Our classifier.} We use Logistic Regression as our  classification engine, which performed better than several others, including SVM, Bayesian networks, and K-nearest-neighbors. 
In \cseed, we use
the Logistic Regression's prediction probability
with a threshold of 0.85 to strike a balance between
sufficient confidence level and adequate number of instances above that threshold.
We found this value to provide better performance
than 0.8 and 0.9, which we also considered.

% with a special focus on the value of our  \cseed approach.

%============================================
%============================================
{\bf A. The IP \idenp problem.}
As we saw in section~\ref{sec:ipDetection}, our classification approach exhibits 98\% precision and 96\% recall on average across all our sites, when we train with ground-truth for each forum.

%\textbf{a.  \idenp: Compare \cseed and \cporting.}

{\bf a. \idenp: 95\% precision with \cseed.}
We show that our \crosstrain approach is effective in
transferring the knowledge between domains.
We use the classifier from \WS and we use it to classify 
 three of the other forums, namely, 
 \OffComm, \ethical, and \HTS.
 Note that we do not include \darkode in this part of
 the evaluation as it did not have sufficient data
 for testing (less than 15  non-address expressions in all its posts).
 
\begin{figure}[t]
    \vspace{-0.5cm}
    \centering
    \begin{minipage}{0.5\textwidth}
        \centering
        \includegraphics[width=\textwidth]{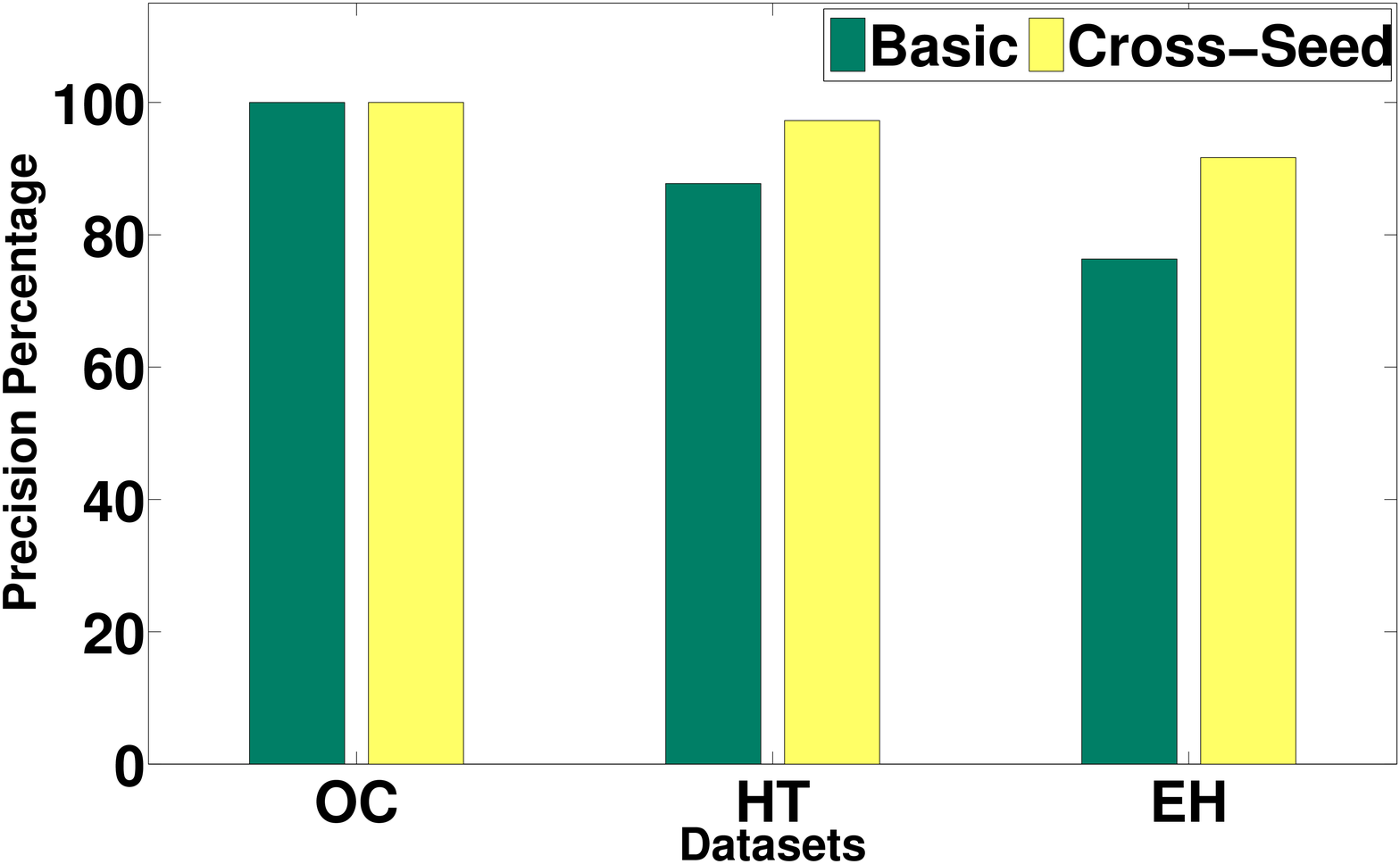} % first figure itself
        {(a)Precision}\label{fig:WSIPPrec}
        \end{minipage}\hfill
    \begin{minipage}{0.5\textwidth}
        \centering
        \includegraphics[width=\textwidth]{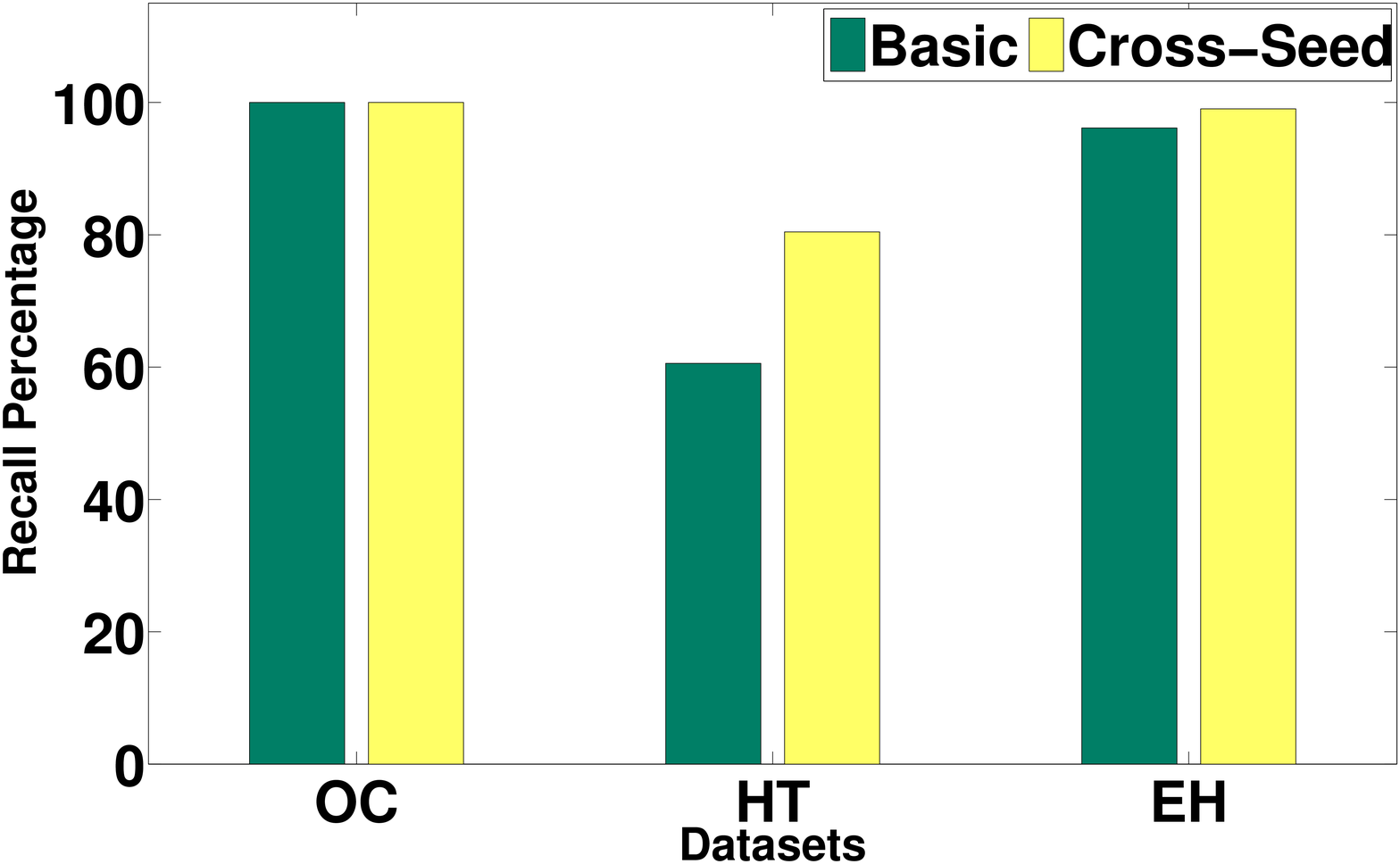} % second figure itself
        {(b)Recall}\label{fig:WSIPRec}
    \end{minipage}
    \caption{\idenp: \cseed improves both Precision and Recall. Using \WS to classify
    \OffComm, \HTS, and \ethical.}
    \label{fig:IPWilder}
    \vspace{-0.5cm}
\end{figure}

In figure~\ref{fig:IPWilder}, we show the results
for precision and recall of \crosstrain
using \cporting and \cseed.
We see that \cseed improves {\it both} precision and recall
significantly. 
For example,  for \HTS,  \cseed increases
the precision from 57\% to 79\% and the recall from  60\% to 78\%.

{\bf b. \idenp: \cseed  outperforms \cporting.} \cseed  improves the  precision by 8\% and recall by 7\% on average for the experiment shown in figure~\ref{fig:IPWilder}.
 The average precision increased from
88\% to 95\% and the average recall increased from 85\% to 97\%.

% {\bf c. \classp: 93\% precision with training data.} We develop  supervised learning algorithm for 
% solving the \classp problem 
% in the case where we have training data for the target forum. Our classifier achieves 
% 93\% precision and 92\% recall on average over all our forums. 

 %------- Michalis stopped here ----------
 % \newcommand{\OffComm}{Offensive Community\xspace}
% \newcommand{\OffCommShort}{Offensive Comm.\xspace}
% \newcommand{\HTS}{Hack This Site\xspace}
% \newcommand{\WS}{Wilders Security\xspace}
% \newcommand{\Ash}{Ashiyane\xspace}
% \newcommand{\WSShort}{Wilders Sec.\xspace}
% \newcommand{\ethical}{Ethical Hackers\xspace}
% \newcommand{\darkode}{Darkode\xspace}

%In Figure \ref{fig:IPWilder},

% On average over all datasets 95\% precision and 93\% recall reported for \WS as the model.
% Selecting \WS dataset as the source and the rest as the target we evaluate the performance of the \cseed method. As depicted in Figure \ref{fig:IPWilder}, even though \cporting works pretty good on the \OffComm dataset but for \HTS and \ethical datasets as the target if the \cseed applies it could significantly improve precision and recall on the point of interest class , i.e. Not-IP class.

%============================================
%============================================
{\bf B. The IP \classp problem.}
We evaluate our approach for solving the \classp problem
without per-forum training data.
As we saw in section \ref{sec:MalIP},  we can achieve 93\% precision and 92\% recall on average across all the forums, when we train with ground-truth for each forum.

\begin{figure}[ht]
\vspace{-0.5cm}
    \centering
    \begin{minipage}{0.5\textwidth}
        \centering
        \includegraphics[width=\textwidth]{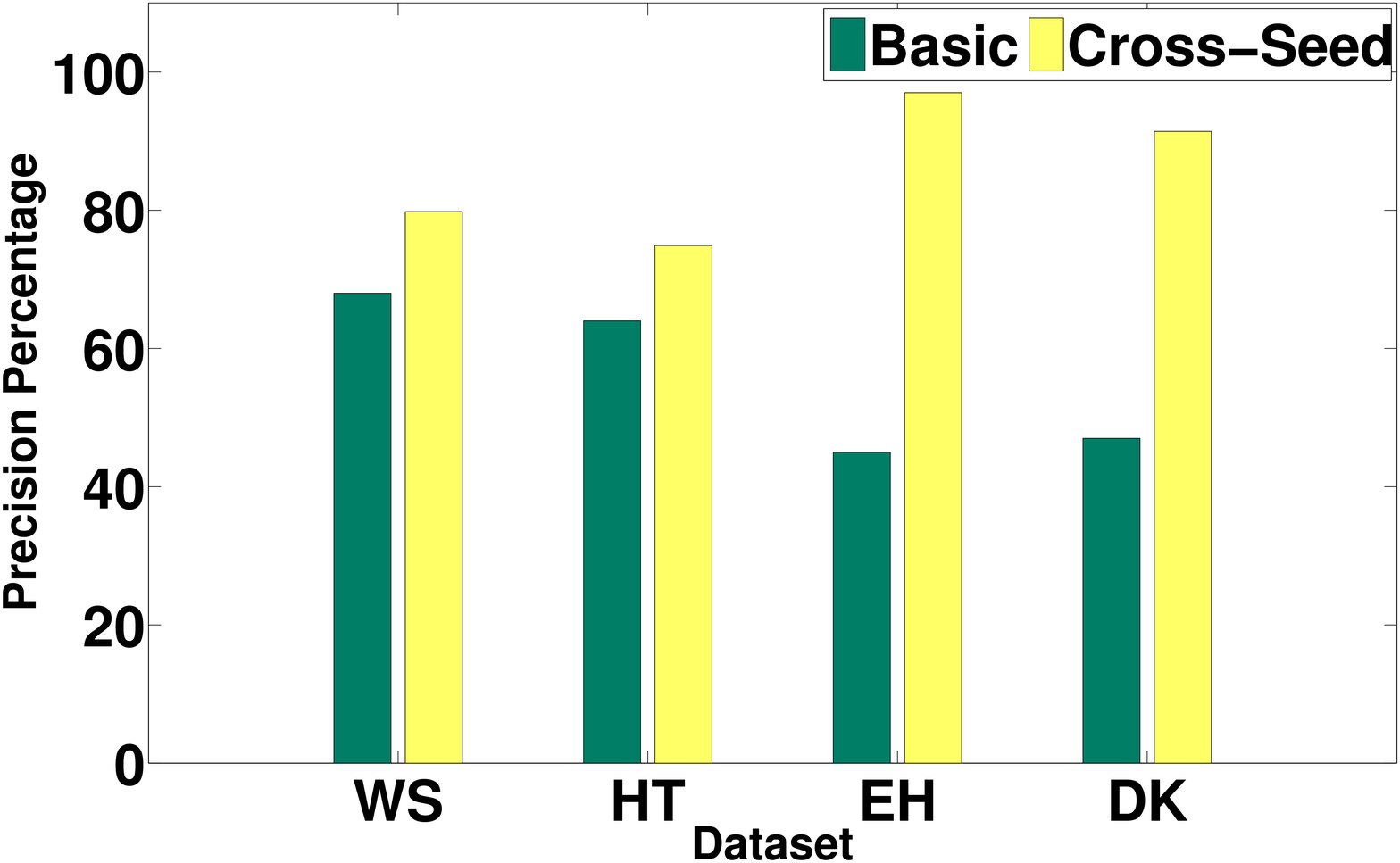} % first figure itself
        {(a)Precision}\label{fig:OffCommPre}
        \end{minipage}\hfill
    \begin{minipage}{0.5\textwidth}
        \centering
        \includegraphics[width=\textwidth]{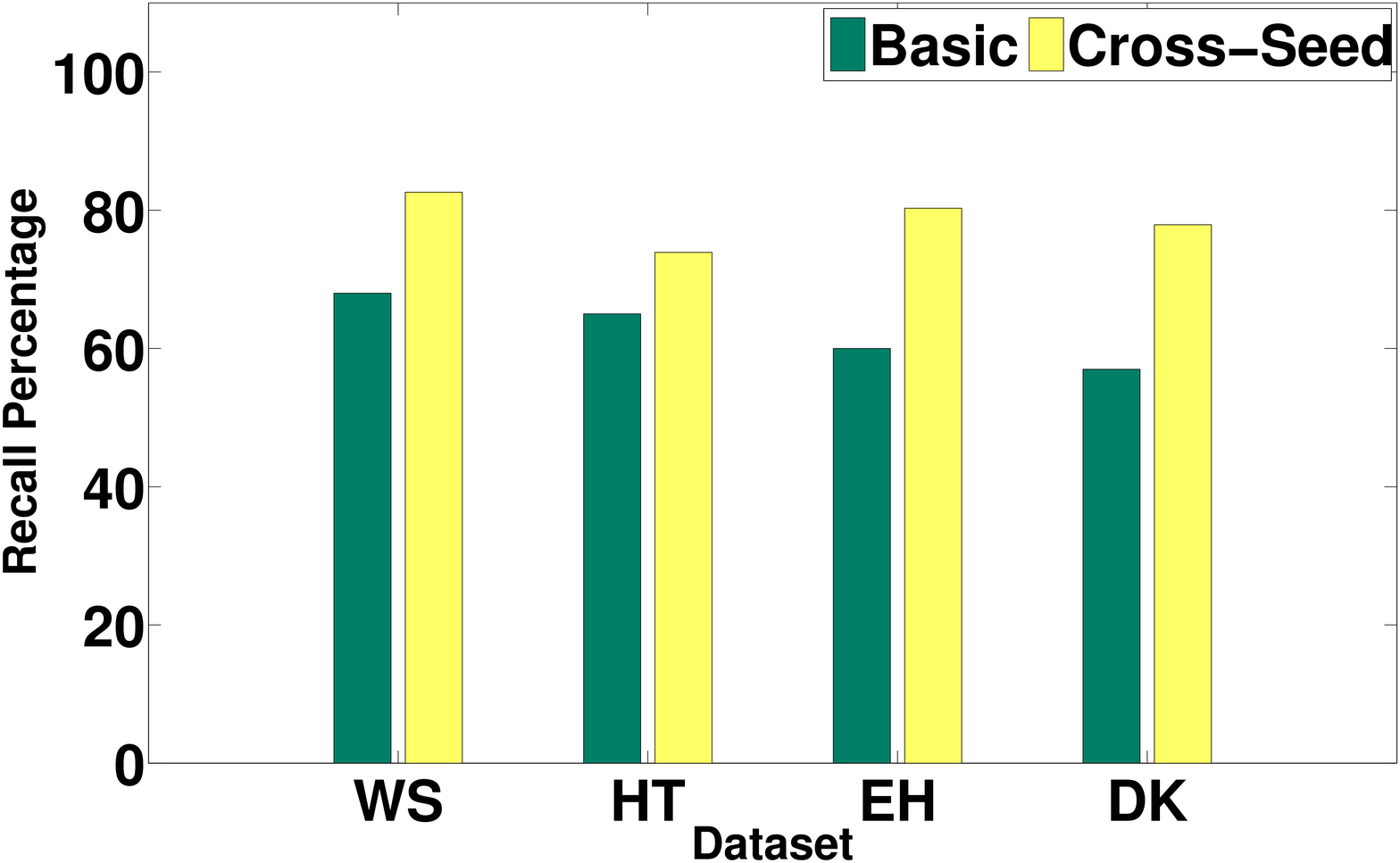} % second figure itself
        {(b)Recall}\label{fig:OffCommRec}
            \end{minipage}
    \caption{
    \classp: \cseed improves both Precision and Recall. Using \OffComm  as source, we classify
    \WS, \HTS, \ethical and \darkode.
    }
    \label{fig:OffCommCompare}
\vspace{-0.5cm}
\end{figure}

{\bf a. \classp: 88\% precision on average with \cseed.} 
Using \OffComm as source, and we classify
    \WS, \HTS, \ethical and \darkode
    as shown in figure~\ref{fig:OffCommCompare}.
Our \cseed approach can provide
88\% precision and 82\% recall on average.
% {fig:OffCommCompare}

%What is improvement over "non intelligent" \crosstrain?

% \begin{figure}
%     \centering
%     \begin{minipage}{0.5\textwidth}
%         \centering
%         \includegraphics[width=\textwidth]{figures/CouplePrecision.eps} % first figure itself
%         \subcaptionbox{(a)Precision}\label{fig:CouplePrec}
%         \end{minipage}\hfill
%     \begin{minipage}{0.5\textwidth}
%         \centering
%         \includegraphics[width=\textwidth]{figures/CoupleRecall.eps} % second figure itself
%         \subcaptionbox{(b)Recall}\label{fig:CoupleRec}
%     \end{minipage}
%     \caption{\classp: Using more source forums improves the precision. Here we use \WS and \ethical as sources,
%     and use \cseed on \OffComm, \HTS, and \darkode.}
%     \label{fig:CoupleTest}
%     \vspace{-5pt}
% \end{figure}

{\bf b. \classp: \cseed  outperforms \cporting.} 
We show
that \cseed  improves the classification compared
to just reusing the classifier from another forum.
In figure~\ref{fig:OffCommCompare},  we show the precision
and recall of the two approaches. Using \OffComm as our source,  we see that  \cseed  improves the
precision by 28\% and recall by 16\% on average
across the  forums compare to the Basic approach.
We also observe that the improvement is
substantial: \cseed improves both precision and recall in all cases.

\begin{table}[ht]
\vspace{-0.5cm}
 \centering 
%\begin{adjustbox}{width=.70\textwidth}
%\small
 \begin{tabular}{|c|c|c|c||c|} 
 \hline
  & \OffCommShort & \HTS & \darkode & Average \\ 
 \hline
 Precision & 3.3 & 20.5  & 17.8  & 13.2 \\ 
 \hline
 Recall & 8.3 & 6.4 & 38.8 &  17.8\\ 
 \hline
\end{tabular}
%\end{adjustbox}
 \caption {\classp: Using two instead of one source forums improves precision and recall on average: Average improvement of using \ethical and \WS as sources together compared to each of them individually.}
 \label{tab:CoupleImprovement}
\end{table}

\vspace{-0.9cm}
{\bf c. Using more source forums improves the \cseed performance significantly.} 
We quantify the effect of having more than one source forums in the classification accuracy of a new forum.
We use \ethical and \WS as our training forums,
and we use \cseed for \OffComm, \HTS, and \darkode.
First, we use the source forums one at a time and 
then both of them together.
In table \ref{tab:CoupleImprovement}, we show the average improvement of having two source forums over having one for each target website. Using two source forums
increases the classification precision by 13\%
and the recall by 17\% on average.

{\bf Discussion: Source forums and training.}
How would we handle a new forum?
Given the above observations, we would currently 
use all our five forums as sources for a new forum.
Overall, we can argue that the more forums we have,
the more we can improve our accuracy.
However, we would like to point out that 
some forums are more ``similar" and thus more
suitable for cross-training. We will investigate
how to best leverage a large group of source forums
once we collect 20-25 more forums.

% We show that, by adding a second source forum, we can improve the precision by 13\% on average over the remaining three forums compared to using one source forum.
% In figure~\ref{fig:CoupleTest},
% %\ref{fig:CouplePrec}.
% we show the results for precision and recall.
% We see that the precision always improves 
% when we use both training forums.

\hide{
---------------

\textbf{1) Compare \cseed and \cporting.}

{\bf d. \classp: 88\% precision with \crosstrain data.} 
We show that our \crosstrain approach can provide
88\% precision and 82\% recall by having \OffComm forums as the source model and test it on four other models. The details of all forums can be seen in the Figure \ref{fig:OffCommCompare}

In order to detect malicious IP addresses we have used \cseed model and compare it with \cporting model in \OffComm forum as the source and the other forums as the target. The results shows, in Figure \ref{fig:OffCommCompare}, the \cseed can improve the performance and outperform \cporting method in all the tested forums. 

{\bf e. Doing \crosstrain  with our  \cseed  approach improves the performance.}

We show
the importance of \cseed method as it increases the precision by 28\% and recall by 16\% on average in \classp. The initialized seeds provides by the model based on \OffComm could increase the performance of the \classp task.

\textbf{2) adding new source forum for training}

{\bf f. Using more source forums improves the \crosstrain performance.}

We show that by adding a second source forum we can improve the precision by 13\% on average over the remaining three forums compared to using one source forum.

As it can be seen in Figure \ref{fig:CoupleTest} By adding more data to the model the precision improve with 8\% of precision in a model of Coupling \WS and \ethical  and testing on the others. In this case we have use the label data of \WS and \ethical forums to train a model based on \cseed method.

}

\vspace{-0.3cm}
\section{Related Work}
\label{sec:related}
\vspace{-0.2cm}

We  summarize  related work clustered into areas of relevance.

{\bf a. Extracting IP addresses from security forums.} 
There two main efforts that focus on IP addresses and security forums \cite{Frank2016,Joobin2017} and 
neither provides the comprehensive solution
that we propose here.
The most relevant work~\cite{Joobin2017} 
does not address the \idenp problem, and sidesteps 
the problem of cross-forum training by assuming
training data for each forum.
The earlier work \cite{Frank2016} focuses on 
the spatiotemporal 
properties of Canadian IP addresses in forums, 
but assumes that all
identified addresses are suspicious and therefore
they did not employ  a classification method,
which is the focus of our work. 

% but 
% seems to not classify the reported IP addresses,
% thus it does not seem to solve the
% \idenp nor the \classp problems.

% Critical Candian system has been studied by extracting the geo-location of the IPs appeared on underground security forums. \cite{Frank2016}.

{\bf b. Extracting other information from security forums.}
Various efforts have attempted to extract other types of information from security forums.  A few recent studies identify malicious services and products in security forums by focusing on their availability and price~\cite{Portnoff2017,Motoyama2011}.

{\bf c. Studying the users and posts in security forums.}
Other efforts study the users of security forums, group them into different classes, and identify their roles and social interactions~\cite{Abbasi2014,Zhang2015,Holt2012,Samtani2015,Shakarian2016}.

{\bf d. Analyzing structured security-related sources.}
There are several studies that automate the extraction of information from structured security documents, extracting ontology and comparing the reported information, such as databases of vulnerabilities, and security reports from the industry~\cite{JonesBHG15,BridgesJIG13,Iannacone2015}.

{\bf Transfer learning methods and applications.}
There is extensive literature on transfer learning ~\cite{daumeiii2007,Dai2007,Dai2007_3} and several good surveys~\cite{pan2010,Weiss2016}, which
inspired our approach.
However, 
to the best of our knowledge, we have not found any
 work that 
 address the same domain-specific challenges
 or uses all the steps of our approach,
which we described in \ref{sec:activeLearning}.
%Our approach was inspired the TrAdaBoost
%We can identify two fundamental types of problems
% There is an extension of AdaBoost algorithm for transfer learning in a set of instance-transfer approach in which they attempt to reweight the source
% domain data to adapt the model with good classified sample. So they are transferring instances from source into destination domain to build a model. In this model source and destination has same feature sets but different marginal distribution. 
% Other work in area can be seen in ~\cite{Dai2007,Dai2007_2,Jiang07}. 
% Feature-based transfer learning in another set of work, in which a common feature representation will be selected to reduce the difference between source and domain feature set ~\cite{Dai2007_3,Ando2005,Blitzer2006} among them a simple feature augmentation method received tons of attention in which the augmented feature set based on for both target and source created, consequently same features set for both train and test domain use for building and testing the model~\cite{daumeiii2007}.

%Active learning is a well-known solution that can be part and strengthen a supervised learning solution~\cite{settles2009active}. Indicatively, we 
%refer to two recent studies that use active learning in the general security space:
%they study  adversarial learning  \cite{Miller2014}, and  detecting outliers in web  and firewall logs \cite{KalyanAI2}, detecting adversarial ads~\cite{Sculley2011}.

\vspace{-0.3cm}
\section{Conclusion}
\label{sec:conclustion}
\vspace{-0.2cm}

 We propose a comprehensive solution for mining
 malicious IP addresses from security forums. 
 A novelty of our approach is  it minimizes the need for human intervention. 
First, once it is initialized with a small number of
security forums, it does not require
 additional training data for each new forum.
To achieve this, we use \cseed, which uses initialization via domain adaptation: we use
a classifier from current forums to create seed information  for the new forum. 
Second, it addresses both the \idenp and \classp problems,
unlike all prior work that we are aware of.
We evaluate our method real data and
we show that: (a) our \cseed approach works
fairly well reaching precision above 85\% on average 
for both classification problems,
(b) \cseed outperforms the \cporting approach,
and (c) using more source forums increases the performance
as one would expect.

Our future plans include: (a) collecting a large number of security forums, (b) exploring the limits of the classification accuracy by using more source forums, 
and (c) exploring additional transfer learning methods.

%conducting a large-scale study of
%the spatiotemporal properties of the identified malicious IP addresses.%\reminder{How about this: Improving the accuracy of the transferred knowledge by using more complex methods }
%\mfal{This is a good idea to do, but I am not sure how easy it is to explain it here in a consise and meaningful way and without sounding as if what we did here is "not clever enough". If you want give it a try and let's see how it reads.}
%\input{070-Acknowledgments}

\bibliographystyle{abbrv}
\bibliography{ref}
\end{document}